\documentclass[useAMS,usenatbib,usefloat]{mnras}
\usepackage{float}
\usepackage{amsmath}
\usepackage{amssymb}
\usepackage{rotating}
\usepackage{tabularx}
\usepackage{gensymb}

\include{journaldefs}
\usepackage[font=footnotesize,labelfont=bf]{caption}
\newcommand{\bhg} {J20395358+4222505}

\newcommand{\vsini} {$v$\,sin\,$i$}
\newcommand{\vmacro} {$\Theta_{\rm RT}$}
\newcommand{\vinf} {$v_\infty$}

\newcommand{\Teff} {$T_{\rm eff}$}
\newcommand{\grav} {log\,{\em g}}
\newcommand{\gravc} {log\,{\em g$_c$}}

\newcommand{\logl}{$\log{L/L_{\odot}}$}
\newcommand{\kms}{$\rm km\,s^{-1}$}

\newcommand{\Rsun} {R$_\odot$}
\newcommand{\Msun} {M$_\odot$}

\newcommand{\lLsun} {log(L/L$_\odot$)}
\newcommand{\Mbol}{M$_{bol}$}
\begin{document}

   \title[The nature of 2MASS \bhg]{\bf The nature of the Cygnus extreme B-supergiant\\ 2MASS \bhg}

\author[Herrero et al.]
  {A. Herrero$^{1,2}$\thanks{E-mail: ahd@iac.es}, S.R. Berlanas$^{3}$, A. Gil de Paz$^{4,5}$, F. Comerón$^6$, J. Puls$^7$,\newauthor
  S. Ramírez Alegría$^8$, M. García$^9$, D.J. Lennon$^{1,2}$, F. Najarro$^9$, S. Simón-Díaz$^{1,2}$, \newauthor
  M.A. Urbaneja$^{10}$, J. Gallego$^{4,5}$, E. Carrasco$^{11}$, J. Iglesias$^{12}$, R. Cedazo$^{13}$,\newauthor
  M.L. García Vargas$^{14}$, \'A. Castillo-Morales$^{4,5}$, S. Pascual$^{4,5}$, N. Cardiel$^{4,5}$,\newauthor A. Pérez-Calpena$^{14}$, P. Gómez-Alvarez$^{14}$, I. Martínez-Delgado$^{14}$
  \\ \\
$^1$Instituto de Astrofísica de Canarias, 38200, La Laguna, Tenerife, Spain\\
$^2$Departamento de Astrofísica, Universidad de La Laguna, 38205, La Laguna, Tenerife, Spain\\
$^3$Departamento de Física Aplicada, Facultad de Ciencias II, Universidad de Alicante,Spain\\
$^4$Departamento de Física de la Tierra y Astrofísica, Universidad Complutense de Madrid, E-28040, Madrid, Spain\\
$^5$Instituto de Física de Partículas y del Cosmos IPARCOS, Fac. de Ciencias Físicas, Univ. Complutense de Madrid, E-28040, Madrid, Spain\\
$^6$European Southern Observatory, Karl Schwarzschild Str. 2, 85748, Garching, Germany\\
$^7$LMU München, Universitätssternwarte, Scheinerstr. 1, 81679 München, Germany\\
$^8$Centro de Astronomía (CITEVA), Universidad de Antofagasta, Av. Angamos 601, Antofagasta, Chile\\
$^9$Centro de Astrobiología, CSIC-INTA, Ctra de Torrejón a Ajalvir km 4, E-28850 Torrejón de Ardoz, Madrid, Spain\\
$^{10}$Institut für Astro- und Teilchenphysik, Universität Innsbruck, Technikerstr. 25/8, 6020 Innsbruck, Austria \\
$^{11}$Instituto Nacional de Astrofísica, Óptica y Electrónica, Luis Enrique Erro No. 1, C.P. 72840, Tonantzintla, Puebla, México\\
$^{12}$Instituto de Astrofísica de Andalucía (CSIC), Apdo. 3004, E-18008, Granada, Spain\\
$^{13}$Departamento de Ingeniería Eléctrica, Electrónica Automática y Física Aplicada, E.T.S. de Ingeniería y Diseño Industrial, Madrid, Spain\\
$^{14}$Fractal, S.L.N.E., Calle Tulipán 2, portal 13, 1A, E-28231 Las Rozas de Madrid, Spain\\
}
\date{Received; accepted }

\maketitle
   
\begin{abstract}
2MASS \bhg~ is an obscured early B supergiant near the massive OB star association Cyg OB2. Despite its bright infrared magnitude (K$_s$= 5.82) it has remained largely ignored because of its dim optical magnitude (B= 16.63, V= 13.68). In a previous paper we classified it as a highly reddened, potentially extremely luminous, early B-type supergiant. We obtained its spectrum in the U, B and R spectral bands during commissioning observations with the instrument MEGARA@GTC. It displays a particularly strong H$_\alpha$ emission for its spectral type, B1 Ia.  The star seems to be in an intermediate phase between super- and hypergiant, a group that it will probably join in the near (astronomical) future. We observe a radial velocity difference between individual observations and determine the stellar parameters, obtaining \Teff= 24\,000 K, \gravc= 2.88$\pm$0.15. The rotational velocity found is large for a B-supergiant, \vsini= 110$\pm$25 \kms. The abundance pattern is consistent with solar, with a mild \ion{C}{} underabundance (based on a single line). Assuming that \bhg~ is at the distance of Cyg OB2 we derive the radius from infrared photometry, finding R= 41.2$\pm$4.0 \Rsun, \lLsun= 5.71$\pm$0.04 and a spectroscopic mass of 46.5$\pm$15.0 \Msun. The clumped mass-loss rate (clumping factor 10) is very high for the spectral type, $\dot{M}$= 2.4$\times$10$^{-6}$ \Msun a$^{-1}$. The high rotational velocity and mass-loss rate place the star at the hot side of the bi-stability jump. Together with the nearly solar CNO abundance pattern, they may also point to evolution in a binary system, \bhg~ being the initial secondary.
\end{abstract}


%
%
\section{Introduction}

Massive stars are a tiny fraction of a given stellar generation. For a Salpeter Initial Mass Function (IMF) we expect only one star with more than 8 solar masses for every 500 stars of smaller mass. Moreover, they are the first members of that generation to evolve and dissapear in only few tens of millions of years, when solar type stars barely begin their life in the Main Sequence. As a consequence, massive stars are scarce objects in the Universe. Their study is additionally complicated by the fact that they are relatively far away (compared to stars of lower mass), tend to be born in binary or multiple systems \citep{sana12,demink13}, and are associated with dense clouds of interstellar material from which they were born or are new stars in the process of formation \citep{motte18}. Thus the number of objects available for detailed observation and analysis is comparatively scant.

Among massive stars, blue hypergiants and extreme supergiants are a rare stage because they have high luminosities and are of very short duration \citep[see e.g.][]{clark12}. Also, they are crucial to understand the evolutionary channels followed by single and multiple massive stars, from O and B stages to Luminous Blue Variables (LBV), Red Supergiants (RSG) and Wolf-Rayets (WR), and ultimately to the end products of massive star evolution: supernovae (SN), neutron stars (NS), black holes (BH), high-mass X-ray binaries, long-duration Gamma Ray Bursts (GRB) and gravitational wave progenitors \citep[e.g.][]{langer12,abbott16}. 

\begin{figure}
   \centering
   \includegraphics[scale=0.40,angle=0]{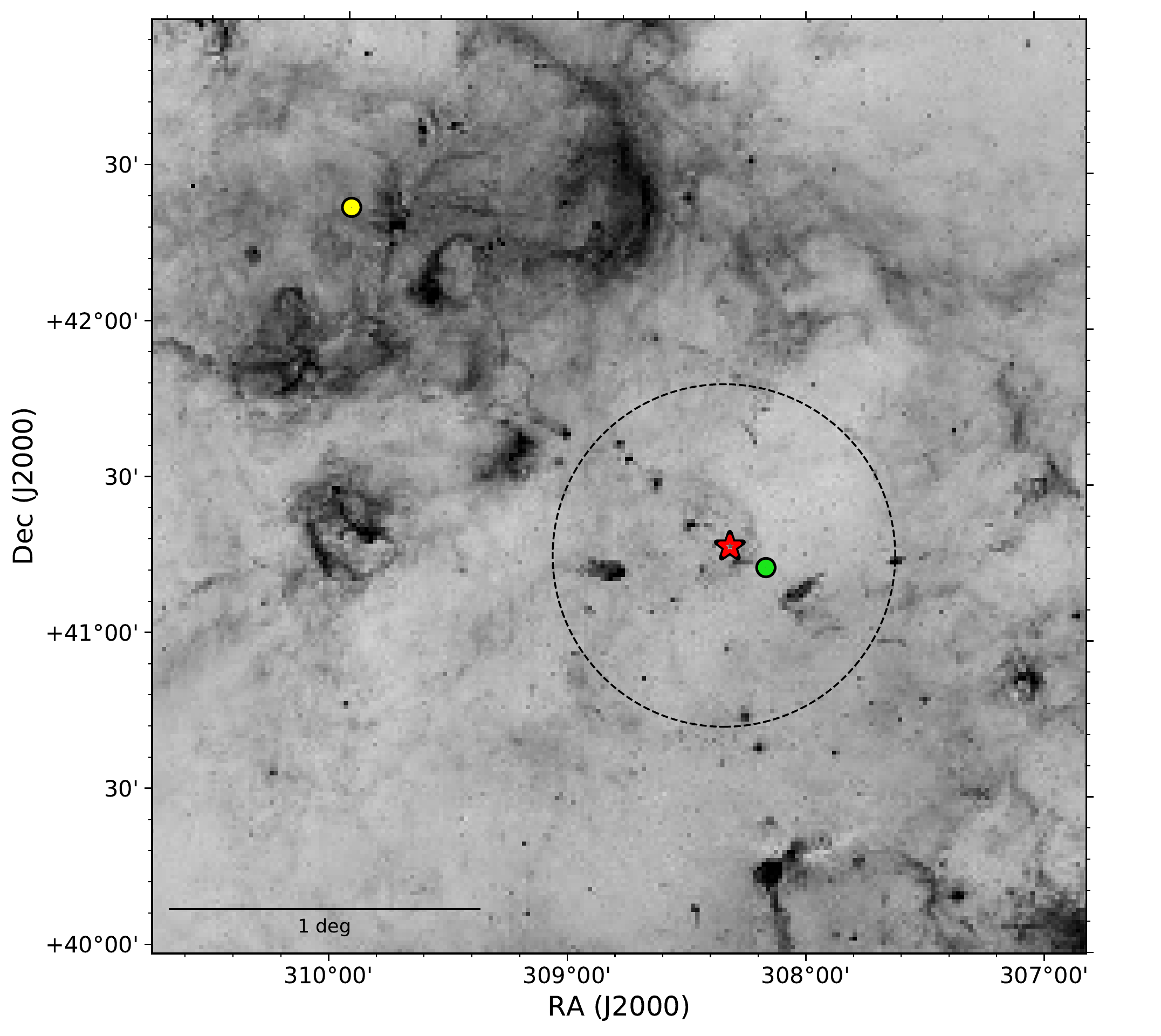}
     \caption{The region of the CygOB2 assocation showing the location of \bhg~ (yellow dot). The red star marks the position of the CygOB2 \#8 multiple stellar system, which may be considered the centre of the association, and the green circle represents CygOB2 \#12. The dashed circle indicates the core of the association as defined by \protect\cite{wright15}
}
         \label{map}
   \end{figure}

Here we report on 2MASS \bhg ~\citep{skrutskie06}\footnote{\cite{comeron12} identify it as  J20395358+4222506. Both entries return the same data in the 2MASS catalog, and it is the only source within 5 arcsec},  a highly reddened object in the vicinity of the CygOB2 association (see Fig.~\ref{map}). It was spectrocopically observed by \cite{berlanas18} and classified as B0 I. These authors quote an apparent magnitude in the B-band of m$_B$= 15.89 \citep{comeron12} and estimate an extinction of A$_v$= 11 for this star or A$_B \approx$ 14.6. Adopting a distance of 1.76 kpc to Cyg OB2 \citep{berlanas18} this implies an absolute magnitude M$_V \approx$ -9.8, or, adopting a B.C.= -2.0,  M$_{bol} \approx$ -11.8 and \logl $\approx$ 6.6\footnote{That value is different from the one obtained in this work because we have used infrared photometry to determine the radius instead of the B-magnitude. Note also that the B-magnitude quoted by \cite{comeron12} from NOMAD is different from the one adopted here from UCAC4. See Sect.~\ref{radius}.}. That would place the star as one of the most luminous objects among the B supergiants, comparable to the B hypergiants listed by \cite{clark12}. In this paper we present new observations of \bhg~ obtained with MEGARA at the 10.4m GTC telescope and a first analysis to obtain its stellar parameters.

\section{Observations}
The observations were carried out with the MEGARA integral-field and multi-object fiber spectrograph \citep{gdp18} during its Commissioning Phase at the Gran Telescopio CANARIAS on the nights of August 29$^{\mathrm{th}}$ and 30$^{\mathrm{th}}$ 2017. The integral-field unit called Large Compact Bundle (LCB hereafter) was used for these observations. The LCB fully covers a field of view of 12.5$\times$11.3\,arcsec$^2$ with a total of 567 hexagonal spaxels of 0.62\,arcsec (diameter of the circle on which the spaxel is inscribed) plus 56 spaxels split in the eight 7-fiber minibundles located at $\sim$2\,arcmin from the LCB centre that are used for accurate simultaneous sky subtraction. \bhg~ was observed with three different Volume Phase Holographic gratings, namely LR-U, LR-B and HR-R. These setups yielded the wavelength coverage specified in Tab.~\ref{obs}. The LR-B setup was observed on August 29$^{\mathrm{th}}$ 2017, while the other two setups were observed on August 30$^{\mathrm{th}}$ 2017. The on-target exposure times and resolving powers achieved are also listed in Tab.~\ref{obs}, together with the signal-to-noise (S/N) ratios. The use of the LCB IFU avoided losing any flux from the star due to small pointing errors (3.5\,arcsec in the case of the LR-B observations on August 29$^{\mathrm{th}}$ 2017) or bad seeing (1\,arcsec on August 29$^{\mathrm{th}}$ 2017 but between 1.5-2\,arcsec on August 30$^{\mathrm{th}}$ 2017).

The data reduction was done using the python-based MEGARA Data Reduction Pipeline \citep{pascual19,cardiel18}. The data processing included the bias subtraction for each of the two amplifiers used for reading out the MEGARA E2V 231-84 CCD, the trimming of the image section of the CCD, tracing the fibers, wavelength calibration, fiber-flat fielding, and flux calibration. The wavelength calibration was performed using ThAr lamps for the LR-U and LR-B setups and ThNe lamps for HR-R. The flux calibration was obtained by observing the spectro-photometric standard stars Feige 15 and BD+40 4032. The processing of the images in each spectral setup was done separately and yielded three different flux-calibrated multi-extension Row-Stacked-Spectra (RSS hereafter) FITS frames that include 623 4300-element flux vectors and positions on the sky of every single fiber.
A faint extended nebulosity is seen in some fibers away from the star, producing faint nebular lines, like \ion{O}{iii}, \ion{S}{ii}, \ion{N}{ii} or H$_\alpha$. However the reduced stellar spectra do not show indications of their presence, nor over- or undersubstraction. 

After data reduction the LR-U and LR-B spectra were degraded to R= 3000 to improve the S/N ratio and all spectra were shifted to the rest frame by correcting for individual radial velocity off-sets. The LR-B spectrum, taken on August 29$^{\mathrm{th}}$ 2017, required a larger correction than the LR-U and HR-R spectra taken the day after (with a difference in radial velocity of about  60 \kms~ between the first spectrum and the other two when correcting them to the rest frame). We tried to confirm the variation by looking at the interstellar medium (ISM) features. The HR-R spectrum, of better quality, shows ISM lines consistent with the stellar lines velocity and with their rest frame, as listed by \cite{herbig95}. The interstellar \ion{Ca}{ii} lines can be identified in the LR-U spectrum with a velocity also consistent with their laboratory wavelength and a small shift of 20$\pm$14 \kms w.r.t. the stellar lines. The situation is more difficult in the LR-B spectrum, that shows a number of strong interstellar features that are however relatively broad. They are again consistent with their rest frame, although with a larger error bar (10$\pm$17 \kms) but show a large shift (68$\pm$23 \kms) with respect to the stellar lines, as we would expect if the star is moving with respect to the ISM; however the ISM radial velocities in LR-B are not fully consistent with those of the HR-R and LR-U spectra after barycentric correction, with a difference of 22$\pm$19 km/s. For this reason, with only one set of spectra for each wavelength range, we refrain from speculating about the properties of the putative binary and will only consider its possible presence when discussing some points in Sec.~\ref{radius}, \ref{photvar} and \ref{evolstat}.

\begin{table*}
\centering
\caption{Details of the observations with the different gratings. Wavelengths are in Angstroms and exposure times in seconds. The S/N ratio in the LR-U grating does not include the bluemost wavelengths, that have a poorer signal}
\label{obs}
\begin{tabular}{lccccc}
\hline
Grating & Central wavelength & Spectral range & Resolving power & Exposure time & S/N \\  
\hline
LR-U & 4025.90 & 3654.32-4391.88 & 5750 & 2700 & 40-55\\
LR-B & 4785.32 & 4332.05-5199.96 & 5000 & 1800 & 40-75 \\
HR-R & 6602.59 & 6405.61-6797.14 & 20050 & 900 & 110 \\
\hline
\end{tabular}
\end{table*}

   \begin{figure*}
   \centering
   \includegraphics[scale=1.0,angle=0]{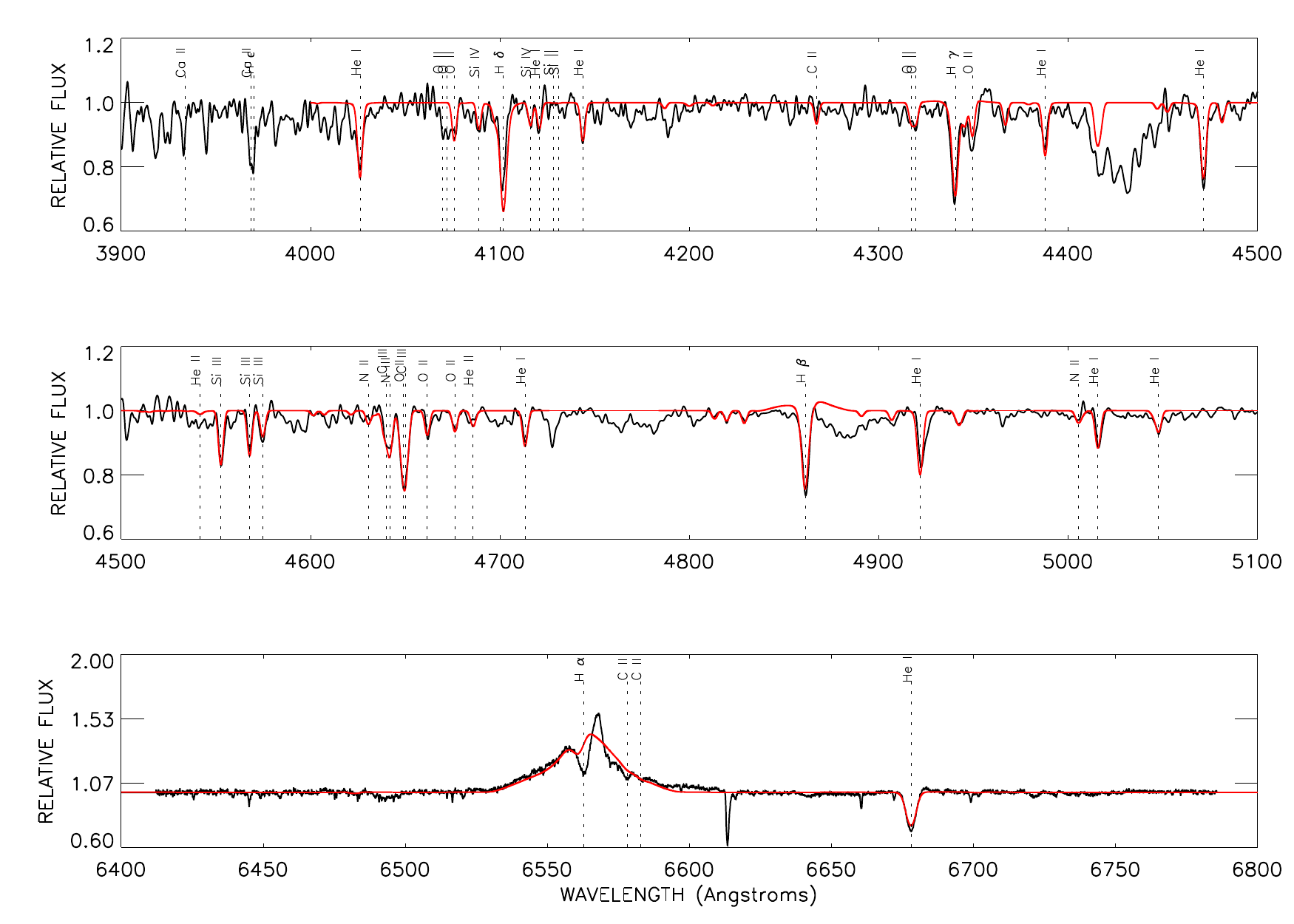}
      \caption{The spectrum of the B hypergiant J20395358+4222505 observed with MEGARA at GTC together with our final fit (see Sect.~\ref{fitparam}. The violet and blue wavelength ranges have been degraded to R=3000, whereas the original R=20050 has been kept for the H$_\alpha$ region. The main lines used in the analysis have been identified for reference.}
         \label{spectrum}
   \end{figure*}
   
   \section{Spectroscopic Analysis}
   \subsection{General description}
   The observed spectrum can be seen in Fig.~\ref{spectrum} together with the final fit described in the next section. In the figure we mark the main spectroscopic features. We identify clearly the Balmer series with the strong emission in H$_\alpha$ and also a very weak red emission peak in H$_\beta$\footnote{The strong diffuse interstellar band redwards of H$_\beta$ could however mask a faint red emission in this case. Note the weak emission in the synthetic spectrum at this point}. The \ion{He}{i} lines are strong and \ion{He}{ii} $\lambda$4686 is weak but marginally present, and probably blended with a weak \ion{Si}{iii} 4683 line. A weak \ion{Mg}{ii} $\lambda$4481 line is also seen. The \ion{Si}{} spectrum is clear with a strong \ion{Si}{iii} triplet at $\lambda\lambda$4552-67-75 and weak \ion{Si}{iv} lines at $\lambda$4089 and $\lambda$4116, the former being blended with an \ion{O}{ii} line. No \ion{Si}{ii} lines could be identified. The \ion{N}{} spectrum displays the \ion{ N}{iii} lines around $\lambda$4640 and for \ion{N}{ii} only the very faint lines at $\lambda\lambda$ 5005-5007 \AA~ could be identified. The \ion{C}{} spectrum is also not very remarkable, although the \ion{C}{ii} lines at $\lambda$6578-83 \AA~ are clearly present. The region around $\lambda$4650 is dominated by the \ion{O}{ii} lines, which shows a rich spectrum.

   We classify the star using the silicon reference frame for the classification of B-type supergiants described by \cite{walborn90}. Specifically the spectral type is defined by the ratio of Si IV 4089 to Si III 4552 that indicates a spectral type of B1. A luminosity class of Ia is indicated by the weakness of the H$_\gamma$ and H$_\beta$ beta lines, hence the star is classified as B1 Ia. In spite of the strong H$_\alpha$ emission, it does not show a clear P-Cygni profile in H$_\beta$, as do all early-B Ia$^+$ stars in \cite{clark12}. Therefore we do not classify it as Ia$^+$, but the star is clearly close to that regime. We suggest that the presence of a P-Cygni profile in H$_\beta$ could be used to distinguish classes Ia and Ia$^+$, presently not clearly separated. This way, a purely spectroscopic criterion would be used, which will strengthen the objectivity in the classification. The presence of HeII 4686 in the spectrum, while it is very weak, is a bit peculiar, although it is well fitted by the synthetic spectrum (see Fig~\ref{spectrum}). It may indicate either helium enrichment or be a result of the high stellar luminosity, as discussed below. 
   
\begin{figure}
   \centering
   \includegraphics[scale=0.5,angle=0]{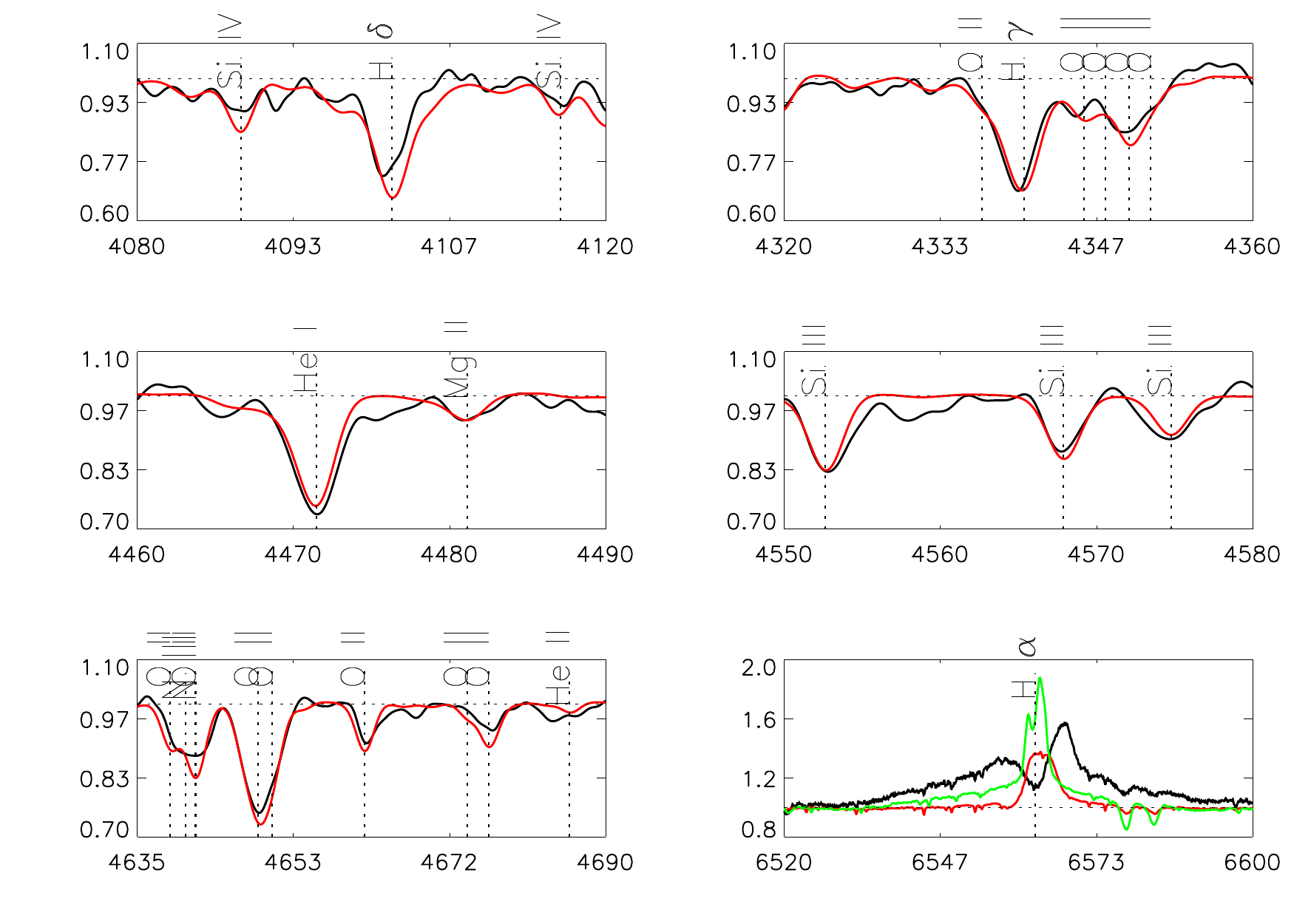}
      \caption{Comparison of the spectrum of J20395358+4222505 (B1 Ia, black) with that of HD\,2905 (BC0.7 Ia, red) and Cyg OB2 \#12 (B3-4 Ia$^+$, green, only in the H$_\alpha$ range) in selected spectral regions. See text for the discussion.}
         \label{spec_comp}
   \end{figure}

   In Fig.~\ref{spec_comp} we compare selected regions of the spectrum of \bhg~ with those of the  BC0.7 Ia supergiant HD\,2905 ($\kappa$ Cas)\footnote{$\kappa$ Cas appears classified as B1 Ia in Simbad \citep{lesh68}. \cite{walborn72} classified it as BC0.7Ia. Because of its \ion{Si}{IV} to \ion{Si}{III} ratio and lack of P-Cygni profile in H$_\beta$ we agree with the classification by Walborn} from the IACOB database \citep{ssimon11, ssimon20} (degraded to the same resolving power as \bhg). HD\,2905 has been recently analyzed by \cite{ssimon18}. The comparison of the \ion{He}{i} 4471\AA~ and \ion{Mg}{ii} 4481\AA~ lines indicates a similar temperature for both stars\footnote{\cite{clark12} indicate that this ratio may depend on the location of the transition region between photosphere and wind. We run some models to test the effect, concluding that this line ratio is not affected in the case of \bhg.}. This is confirmed by the behaviour of the \ion{Si}{iii} lines at 4552-72\AA. The \ion{Si}{iv} lines around H$_\delta$ are slightly weaker in the case of \bhg, whereas \ion{He}{ii} 4686\AA, weak in HD\,2905, is marginally present in \bhg. These differences can be partly explained considering the different H$_\alpha$ emission of both stars, pointing to a stronger or denser stellar wind or to a slightly enriched \ion{He}{} spectrum in \bhg, or both. The \ion{N}{ii} spectrum (particularly the line at $\lambda$4630) is similar in both stars, indicating that \bhg~ could also be of BC-type, but we will wait for a better spectrum to confirm this nuance.  The behaviour of H$_\delta$ and H$_\gamma$ is not completely consistent, perhaps affected by the strong wind producing the H$_\alpha$ emission or by the noise in the region of H$_\delta$, but they point to a similar gravity. We also show in the figure the H$_\alpha$ emission of the hypergiant CygOB2 \#12 (B3-4 Ia$^+$) obtained by A. Pellerin with the High Resolution Spectrograph at the Hobby-Eberly Telescope (again degraded to compare with \bhg; an analysis of Cyg OB2 \#12 and other blue hypergiants can be found in \cite{clark12}). This illustrates that the H$_\alpha$ emission in \bhg~ is indeed very strong and broad. From this comparison, we expect the stellar parameters of \bhg~ to be similar to those of HD\,2905, but with a stronger stellar wind.

   \subsection{Stellar parameters}
   \label{fitparam}
   
   We have determined the projected rotational velocity and the extra broadening (so-called {\it macroturbulence}, $\Theta$) from \ion{Si}{iii} $\lambda$4552 \AA~ (at R=3000 and in the original R=6000 spectrum), \ion{He}{i} 4713 \AA~ (at R= 3000) and \ion{He}{i} 6678 \AA~ (at R= 20050) \footnote{In spite of the Stark broadening, the low gravity combined with the high vsini render \ion{He}{i} lines useful for this purpose} using the {\sc iacob-broad} package \citep[see][]{SH14}, which uses both a Fourier transform and a goodness-of-fit technique to determine these values, assuming a classical rotational and a radial-tangential profile \citep[see][]{gray08}. 

   All lines gave consistent results, and we obtained \vsini= 110$\pm$25 km s$^{-1}$ and \vmacro= 86$\pm$30 km s$^{-1}$.
   This is a large rotational velocity for a B supergiant, doubling that of HD\,2905 \citep{ssimon18} or the highest one in the sample of \cite{clark12} (BP Cru, 55 km s$^{-1}$). We may speculate that the high mass-loss rate found in Sect.~\ref{wind} should be a recent phenomenon, or otherwise the star should have slowed down. Alternatively, it may indicate an efficient transport of angular momentum from the interior to the surface.
   
   We have used the atmosphere code FASTWIND \citep{santolaya97, puls05, rivero11} to estimate the stellar parameters of J20395358+4222505. FASTWIND calculates the emergent stellar spectrum in spherical geometry with mass-loss and NLTE conditions. Because of the relatively low S/N ratio of our spectrum in the blue-violet region and the large parameter space we carry out an estimation of the stellar parameters looking by eye for the model that best fits all diagnostics symultaneously, rather than using our semi-automatic tools. Uncertainties are estimated by varying the stellar parameters until we find that the fit is not acceptable. A more accurate determination of stellar parameters and errors will be carried out in the future when higher S/N ratio spectra are available.
   
   The effective temperature is mainly constrained by the \ion{Si}{iii-iv} and \ion{He}{i-ii} balances and the \ion{He}{i} 4471 to \ion{Mg}{ii} 4481 ratio. The gravity is mainly derived from the fit to the wings of the higher Balmer lines (H$_\gamma$ and H$_\delta$; H$_\epsilon$ is not used because of the low S/N ratio in the region). After the comparison with HD\,2905 we give slightly more weight to H$_\gamma$. The mass-loss rate is obtained from H$_\alpha$ and H$_\beta$. 
   The strong emission in H$_\alpha$ allows us to exceptionally constrain the terminal wind velocity and the exponent of the wind velocity law $\beta$ from the optical spectrum, and thus we obtain v$_\infty$= 1500$\pm$200 km s$^{-1}$ and $\beta$= 1.2$\pm$0.1. The microturbulence has been set to 20 km s$^{-1}$ under the initial assumption that the silicon abundance is solar. This value is close to the one obtained by \cite{ssimon18} of 19 km s$^{-1}$ for HD\,2905. These authors do not consider clumping, and thus we have used the analysis of blue hypergiants made by \cite{clark12} as a reference. We have adopted a clumping parameter\footnote{Note that FASTWIND and CMFGEN, the code used by \cite{clark12}, characterize clumping in different ways: either in the form of a clumping factor (FASTWIND, f$_{\rm cl}= {<\rho^2>\over{<\rho>^2}}$ ) or a volume filling factor (CMFGEN), which under typical assumptions are inverse to each other, f$_{\rm cl}= $f$_{\rm v}^{-1}$ as it has been assumed above when giving Clark et al. numbers} f$_{\rm cl}$=10, intermediate between the values given by \cite{clark12} for Cyg OB2\#12 (f$_{\rm cl}$= 25) and HD 190603 (f$_{\rm cl}$= 4). Thus our mass-loss rate will be equivalent to an unclumped mass-loss rate higher by about a factor of $\sqrt{10}$. We have adopted a linear clumping law, increasing from f$_{\rm cl}$=1 (unclumped) at $v_1= 0.2 v_\infty$ to f$_{\rm cl}$= 10 at $v_2= 0.5  v_\infty$, and keeping it constant at this value until $v_\infty$ This choice has been driven by the need to get a high emission in H$_\alpha$ while not getting too much in H$_\beta$ and H$_\gamma$. Varying these parameters modifies the quality of the fit, but does not change the other stellar parameters beyond the adopted errors.

      Final stellar parameters resulting from the spectral fit are given in Tab.~\ref{params}. Stellar radius, luminosity, mass and mass-loss rate require the distance and extinction determination, that will be described in Sect.~\ref{distance} and~\ref{radius}. The final adopted fit is shown in Fig.~\ref{spectrum} (calculated with the abundances given in Sect.~\ref{abun_sec}). Spectral features are well reproduced, except H$_\alpha$ in the core and, to a lesser extent, H$_\beta$ in the wings, both very sensitive to the details of the adopted clumping law. In particular, we carried out model calculations that indicate that the central absorption in the core of H$_\alpha$ varies as a consequence of the clumping depth dependence. The core of H$_\delta$ is also not well reproduced, but H$_\gamma$, with better S/N ratio, shows a very good fit. 

   \subsection{Abundances}
   \label{abun_sec}

   Abundances are estimated by fitting the different ions in the observed spectrum. Microturbulence is again fixed at 20 \kms~ for all elements, the same one adopted for the Si ionization balance. All abundances are given by number and were initially adopted to be solar and then modified to reach a better fit to the observed spectrum. Most \ion{He}{} lines are well fitted with a solar abundance, but the strong \ion{He}{i} 4471 \AA~ line points to a slightly enriched He abundance, which is consistent with the detection of \ion{He}{ii} 4686. Thus we adopt $\epsilon= \frac{N(He)}{N(H)+N(He)}=$0.12$\pm$0.04, a mildly enriched \ion{He}{} abundance, but still consistent with the solar one. The N abundance is poorly constrained, as the \ion{N}{iii} lines around 4630-4640 are blended with O and C and are highly sensitive to wind details \cite[see][]{rivero11}, whereas the \ion{N}{ii} lines from 5666 to 5710, clear in this spectral type, are outside the spectral range of our observations. Thus we have to rely on the very weak \ion{N}{ii}  5005.15 and 5007.33 \AA~ lines (a blend in our spectrum). These lines are consistent with a solar N abundance. The absence of \ion{N}{ii} 4630.54 and \ion{N}{iii} 4634.13 (that should be present if the N abundance is higher than solar) is also consistent with a solar N abundance. Taking into account the uncertainties in stellar parameters, we determine a value of log(N/H)+12= 7.83$\pm0.15$ (with 7.83$\pm$0.05 being the solar value from \citealt{asplund09}). The abundance of \ion{C}{} relies on the \ion{C}{ii} $\lambda$4267 line, which is weak. Unfortunately, the \ion{C}{ii} 6578-83 lines are not well reproduced by our model. We estimate a \ion{C}{} abundance of log(C/H)+12= 8.13$^{+0.20}_{-0.25}$, a factor of 2 below solar, where errors have again been obtained by fitting the line with varying stellar parameters. However, given the uncertainties, the line weakness and the moderate S/N in the region we cannot completely rule out a solar abundance. For Mg we have only the doublet line at $\lambda$4481, that fits at a slightly lower than solar abundance, but fully compatible with solar (we obtain log(Mg/H)+12= 7.53$\pm$0.15), where the relatively low uncertainty is probably an understimate as we have only one line available. On the contrary, for O and Si we have several lines and thus the same uncertainty of 0.15 dex is more representative. We obtain in both cases solar abundances, with uncertainties determined by varying the stellar parameters (except the microturbulence, that we kept fixed through the whole analysis). Abundances are given in Tab.~\ref{abun_tab}. 

\begin{table}
\centering
\caption{Stellar parameters for \bhg. The stellar mass has been calculated with the gravity corrected for centrifugal acceleration (note that the radius enters in this correction). The whole set of values is given for the adopted distance, together with the statistical errors of the analysis. For parameters depending on the distance, we give the values corresponding to the extreme lower and upper distances. Similar statistical errors would apply to them. No error is given for the adopted value of microturbulence. The wind parameter Q= $\dot{M}\over{(v_\infty R)^{3/2}}$ and the Modified Wind Momentum, D$_{mom}$= $\dot{M}$ \vinf (R$_*$/R$_\odot$)$^{0.5}$, are given in c.g.s. units. The latter will be discussed in Sect.~\ref{wind}. See text for other details.}
\label{params}
\begin{tabular}{lccc}
  \hline
  Parameter  & adopted distance & range of distances \\
                  &      (1.76 kpc)       &     (1.5 -- 2.3 kpc)     \\
  \hline
  \vsini (\kms) & 110$\pm$25& \\
  \vmacro (\kms) & 86$\pm$30 & \\
  \Teff (K) & 24\,000$\pm$500 & \\
  \grav (dex cgs) & 2.85$\pm$0.15 & \\
  \gravc (dex cgs) & 2.88$\pm$0.15 & 2.88 -- 2.87 \\
  logQ (dex cgs) & -12.81$\pm$0.10 & \\
  $\xi $ (\kms) & 20. & \\
  $\epsilon= \frac{N(He)}{N(H)+N(He)}$ & 0.12$\pm0.04$ & \\
  $\beta$ & 1.2$\pm$0.1 & \\
  \vinf (\kms) & 1500$\pm$200 & \\
  R/R$_\odot$ & 41.2$\pm$4.0 & 34.6 -- 53.5 \\
  log(L/L$_\odot$) & 5.71$\pm$0.04 & 5.55 -- 5.93 \\
  M$_{bol}$ (mag) & -9.52$\pm$0.10 & -9.14 -- -10.09 \\
  M/M$_\odot$ & 46.5$\pm$15.0 & 32.9 -- 77.4 \\
  $\dot{M} \times 10^6$ (M$_\odot$ a$^{-1}$) & 2.40$^{+0.20}_{-0.30}$ & 1.8 --3.6 \\
  f$_{\rm cl}$ & 10. & \\
  $v_1/v_\infty$  &  0.2 & \\
  $v_2/v_\infty$  &  0.5 & \\
  log (D$_{mom}$) (cgs) & 29.16$\pm$0.11 & 29.01 -- 29.39 \\
  \hline
\end{tabular}
\end{table}

\begin{table}
\centering
\caption{Abundance numbers for \bhg. For reference we also give the solar values by \protect\cite{asplund09}}
\label{abun_tab}
\begin{tabular}{ccc}
\hline
 Element & log(X/H)+12 & Solar \\
 \hline
  He & 11.08$\pm$0.14 & 10.92$\pm$0.01 \\
  C &  8.13$^{+0.20}_{-0.25}$ & 8.43$\pm$0.05 \\
  N & 7.83$\pm$0.15 & 7.83$\pm$0.05 \\
  O & 8.69$\pm$0.15 & 8.69$\pm$0.05 \\
  Mg & 7.53$\pm$0.15 & 7.60$\pm$0.04 \\ 
  Si & 7.51$\pm$0.10 & 7.51$\pm$0.03 \\
  \hline
\end{tabular}
\end{table}

\section{Discussion}

  \subsection{Distance} 
  \label{distance}
  The distance to \bhg~ is not well determined.
  Based on Gaia Data Release 2 (DR2) measurements \cite{bailer18} obtain a distance of 650 pc from its parallax of 1.54$\pm$0.25 mas, whereas eDR3 gives a parallax of 0.439$\pm$0.09 mas after correcting from zero point \citep{lindegren21}, which would correspond to a plain distance of 2.28 kpc. However, both these measurements have a large Renormalized Unit Weight Error (RUWE, see \citealt{lindegren18}) that amounts to 2.88 in the first case and to 2.95 in the second. This is more than twice the maximum recommended value of 1.4. The short distance is difficult to reconcile with the high extinction towards this object\footnote{Interestingly, a similar situation arose in DR2 for Cyg OB2\#12, the blue hypergiant in the region \citep[see][]{berlanas18} but it has been solved in eDR3}, its stellar parameters and its strong H$_\alpha$ emission (implying a high luminosity) and we discard it. The position of \bhg~ in the sky coincides with the star-forming region DR21, located at 1.5 kpc \citep{rygl12}. Given the strong extinction towards \bhg~ we adopt this as a lower limit for its distance. To adopt an upper limit we rely on the recent work by \cite{pantaleoni21}. These authors have presented a spatial distribution of the massive stars in the solar neighbourhood. In their Fig. 5 we clearly identify the stellar clustering that corresponds to the Cygnus X region and extends from 1.5 to 2.0 kpc. Beyond that distance the stellar density decreases significantly (the Cygnus arm ends). Therefore we adopt the distance of 2.3 kpc given by eDR3 as an upper limit, but note that there is a small probability that the star is at an even greater distance.

\begin{figure}
   \centering
   \includegraphics[scale=0.4,angle=0]{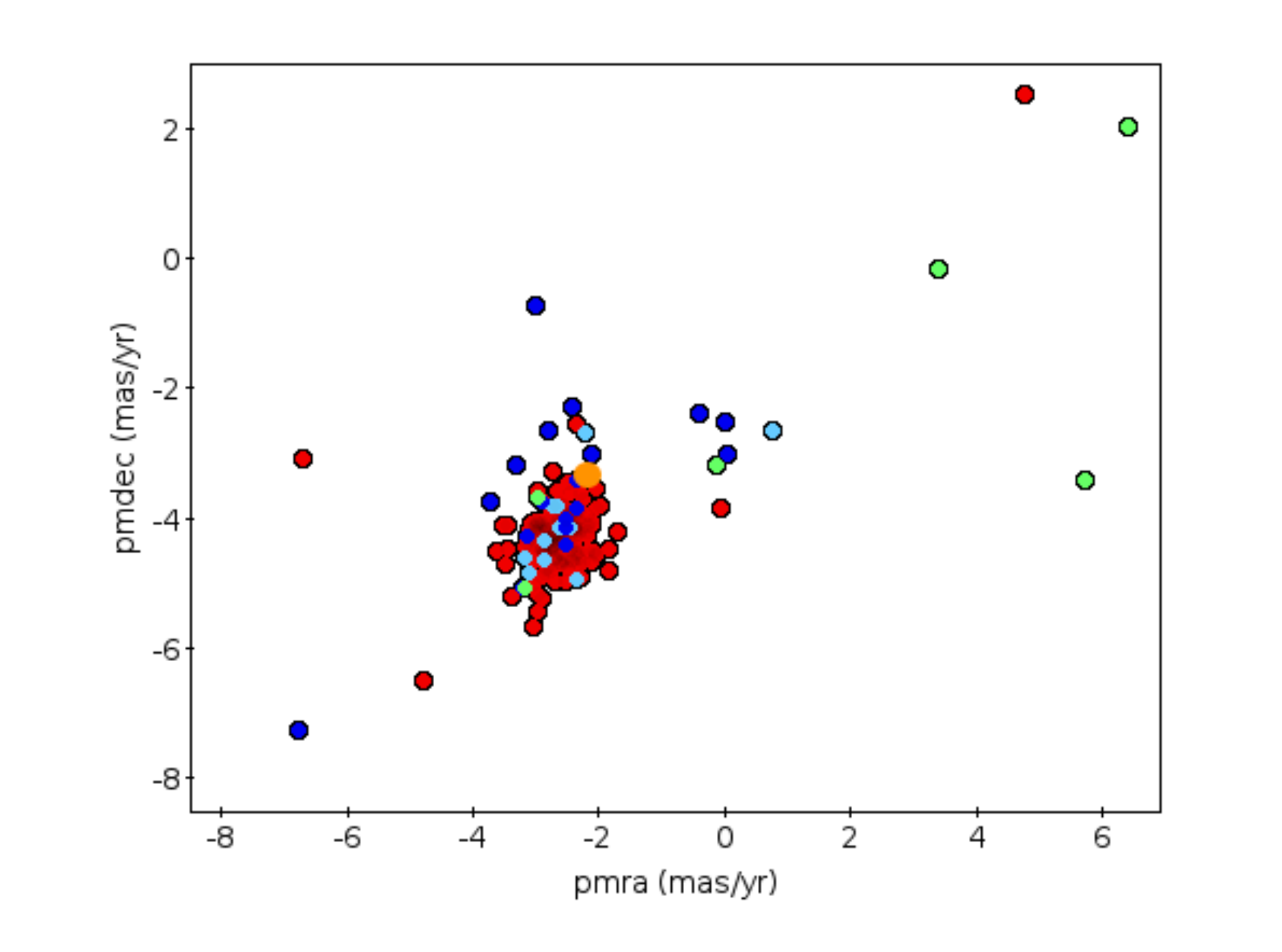}
      \caption{Gaia eDR3 proper motions of \bhg~ (orange circle) and OB-type Cyg OB2 stars. The different colours identify the groups found by \protect\cite{berlanas19}: group 1 (at 1.35 kpc, blue circles); group 2 (at 1.76 kpc, red circles), group 0 (between group 1 and 2, cyan circles) and group 3 (fore- or background stars, green circles)}
         \label{propmot}
       \end{figure}

  The Cyg OB2 association lies very close in the sky (slightly more than 1$\degree$, see Fig.~\ref{map}) and a simple hypothesis would be to assume that the star belongs to the surroundings of this association and is at the same distance. This assumption is supported by the peculiar spectral type, B1 Ia, that we would expect to be associated to a massive star-forming region. However, the star is quite isolated, and its proper motion, ($\mu_\alpha$= -2.139$\pm$0.101 mas a$^{-1}$, $\mu_\delta$= -3.351$\pm$0.116 mas a$^{-1}$, eDR3 data) is similar to that of the main stellar group in Cyg OB2\footnote{DR2 data give $<\mu_\alpha>$= -2.109$\pm$0.451 mas a$^{-1}$ and $<\mu_\delta>$= 0.073$\pm$0.533 mas a$^{-1}$ \citep{berlanas20}} and other stars in the region (see Fig.~\ref{propmot}), suggesting that it is not a runaway, which makes its connection to Cyg OB2 less plausible. There is a possibility that the star is at a larger distance than Cyg OB2.
  
  The distance to Cyg OB2 has been determined by different authors. \cite{massey91} and \cite{torres91} place the association at 1.8 kpc based on stellar photometry and spectroscopy. Later, \cite{hanson03} places it at a shorter distance of 1.5 kpc, that \cite{negueruela08} find consistent with their data\footnote{\cite{clark12} prefer the distance of 1.75 kpc for Cyg OB2 \#12 in their analysis}. More recently, \cite{rygl12} find a similar distance of 1.4 kpc on the basis of the radio parallaxes to four masers in star-forming regions close to Cyg OB2, thus indicating that this may be the distance to the whole Cygnus X region. More recently, again based on Gaia DR2 data, \cite{berlanas19} have found the existence of two groups of OB stars in Cyg OB2, a nearby one at 1.35 kpc formed by a small number of relatively dispersed objects and a second one at 1.76 kpc that includes most of the stars and can be identified with the OB association and that is consistent with the larger distance modulus found by \cite{massey91} and \cite{torres91}. This value is also consistent with that found by \cite{maiz20} for the two main clusters that form the main group of Cyg OB2 (the Villafranca O-007 and O-008 -- also identified by \cite{bica03}-- around Cyg OB2\#22 and Cyg OB2\#8 respectively).Thus we adopt 1755$^{+23}_{-19}$ pc from \cite{berlanas18} as the distance to Cyg OB2 and \bhg, where for the error we have adopted their statistical errors. As lower and upper limits in distance we adopt the values of 1.5 and 2.3 kpc, respectively. 
  
  \subsection{Extinction and stellar parameters} 
  \label{radius}
       Given the adopted distance we can determine the radius taking into account the observed photometry and the extinction. We take the B and V magnitudes from the UCAC4 catalogue \citep{zacharias13} and the J, H and K$_s$ magnitudes from the 2MASS catalogue \citep{cutri03}. We adopt the extinction law by \cite{maiz14} (M14), derived for 30 Dor but applicable to Galactic obscured systems, as the same authors have shown. The M14 law leaves the R$_{5495}$ value free (similar but not identical to R$_V$), which allows to modify the extinction ratio between different bands. We point out here as a cautionary remark that this extinction law, although using a more extended range of reddening values than any other one, has been derived for extinctions smaller than that of \bhg.

       With the extinction law, and using the magnitudes that come from our final model atmosphere after integrating the emergent flux distribution over the appropriate photometric filters, we can derive the stellar radius and total extinction in any photometric band. We have used the infrared magnitudes to determine the radius, as these are less affected by extinction and are quite independent of the R$_{5495}$ value adopted. We use then the combined optical and infrared photometry to get R$_{5495}$. For a typical value of R$_{5495}$= 3.1 we find strong differences between the radii derived from the optical (B, V) and infrared (J, H, K$_s$) bands. We find a consistent radius from these bands for R$_{5495}$= 2.8$^{+0.2}_{-0.1}$, close to the value given by \cite{wright15} (R$_V$= 2.9). With these values  we obtain those quoted in Tab.~\ref{photometry} for the extinction and find a radius R= 41.2$\pm$4.0 \Rsun. From here we derive a luminosity \lLsun= 5.71$\pm$0.04, a bolometric magnitude \Mbol= -9.52$\pm$0.10 and a present stellar mass M= 46.5$\pm$15 \Msun, where the stellar mass has been calculated from the gravity corrected from centrifugal acceleration. The error in the radius, and thus the errors of luminosity, mass and other physical magnitudes, is mainly determined by the inconsistency between the optical and the infrared photometry and, as provided here, not by the adopted uncertainty in the distance. The values corresponding to the lower and upper extreme distances are R= 34.6 \Rsun, \logl= 5.55 and M= 32.9 \Msun (for 1.5 kpc) and R= 53.5 \Rsun, \logl= 5.93 and M= 77.4 \Msun (for 2.3 kpc).

       Could a putative companion to \bhg~ affect our determination of stellar parameters? We have chosen HD\,2905 to simulate the overlapping of \bhg~ with other stars using spectra from the IACOB database. With the S/N and resolving power of our observations we find that at least a conservative flux ratio of 1:0.25 is required for the diagnostic lines to start to be affected. Given the large radius and visual intrinsic brightness of \bhg~ this excludes the possibility of spectral contamination by a hot stripped star or even a hot luminosity class V star, that should be of a spectral type later than O4V (the type of \bhg~ at the Zero Age Main Sequence). An evolved secondary should be at least as bright as an O5 III star, but descended from a star less massive (at ZAMS) than \bhg, which would imply previous mass-transfer and a higher initial mass for \bhg. Even in this case, the change in the parameters derived from the spectrum would be very small and the effect on radius, mass and luminosity through the change in magnitude would remain within the limits given in Table~\ref{params}. There is however the possibility of a spectroscopically very similar companion, that would not modify the spectroscopic parameters, but would imply a significant change in magnitude ($\approx$0.7 mag.). The radius would be 30 \Rsun, with \logl= 5.43 and M= 25.1 \Msun. The fact that we derive normal abundances for all analyzed elements but carbon is also a hint of small, if any, contamination in the spectrum (see Sect. 4 in \citealt{lennon21}) except again for the case of two similar stars.

  \subsection{Photometric variability} 
  \label{photvar}

\begin{figure}
   \centering
   \includegraphics[scale=0.5,angle=0]{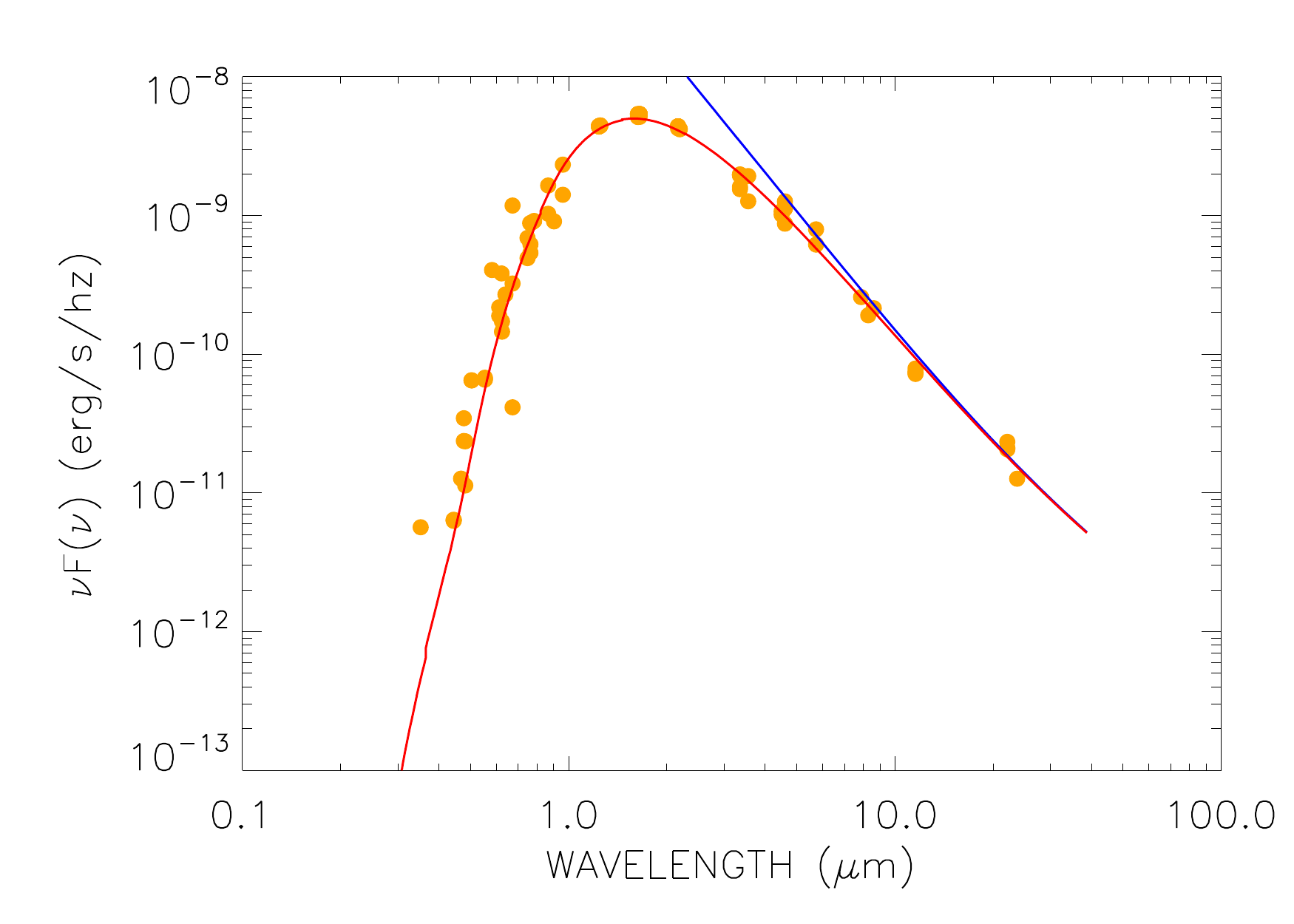}
      \caption{Comparison of the observed photometry from the VizieR Service Photometry Viewer Tool (orange dots; references for the data are given therein) with the continuum flux of our final model, scaled for the adopted distance and corrected from extinction (red line). Because of the hot temperature, only the Balmer jump is clear when plotting the continuum flux in the wavelength region shown . The Paschen jump is present but hard to see in the figure. The blue line is the model flux without the extinction correction, it peaks at around 1100 \AA~ and reaches 6.2$\times$10$^{-6}$ erg s$^{-1}$ Hz$^{-1}$. See text for details.}
         \label{sed}
   \end{figure}

       We have compared our model with the existing photometry in the range 0.3-25 $\mu$m. Fig.~\ref{sed} shows the photometric data from the VizieR Service Photometry Viewer Tool (developed by A.-C. Simon and T. Boch; see also \cite{ochsenbein00}), compared to our final model flux with the derived radius, scaled for the adopted distance to Cyg OB2 and \bhg~ and corrected for extinction using the M14 law. We see that the overall agreement is excellent, in spite of the relatively large dispersion in the observed data. This may point towards source variability, although the large extinction, particularly at shorter wavelengths, introduce additional difficulties in the mesurements. The good agreement between the model and the observation at larger wavelengths (and the absence of any significant difference with the extinction uncorrected model beyond 10 $\mu$m) indicate that there is no emission excess at longer wavelengths due to circumstellar material.

  \begin{table*}
\centering
\caption{Photometric and extinction values for \bhg. Photometry is from the UCAC4 and 2MASS catalogues. \protect\cite{maiz14} has been used for the extinction law}
\label{photometry}
\begin{tabular}{ccccccccc}
\hline
 B & V & J & Ks & R$_{5495}$ & A$_B$ & A$_V$ & A$_J$ & A$_{Ks}$   \\
 \hline
16.63 & 13.682 & 7.345$\pm$0.029 & 5.822$\pm$0.017 & 2.8$^{+0.2}_{-0.1}$ & 12.8  & 9.7 & 2.8 & 1.1 \\
\hline
\end{tabular}
\end{table*}

       As mentioned before, the B-magnitude given by \cite{comeron12} from the NOMAD catalog \citep{zacharias04} and the one used here from the UCAC4 one do not agree. Therefore we have searched the databases for variability. No information could yet be extracted from Gaia, as the object is not included in the list of targets for which epoch information is available. The information in 2MASS is also very scarce, with only two measurements the same day with identical values within errors. PanStarrs provides also a few measurements, but nothing can be concluded from them.

\begin{figure}
   \centering
   \includegraphics[scale=0.35,angle=270]{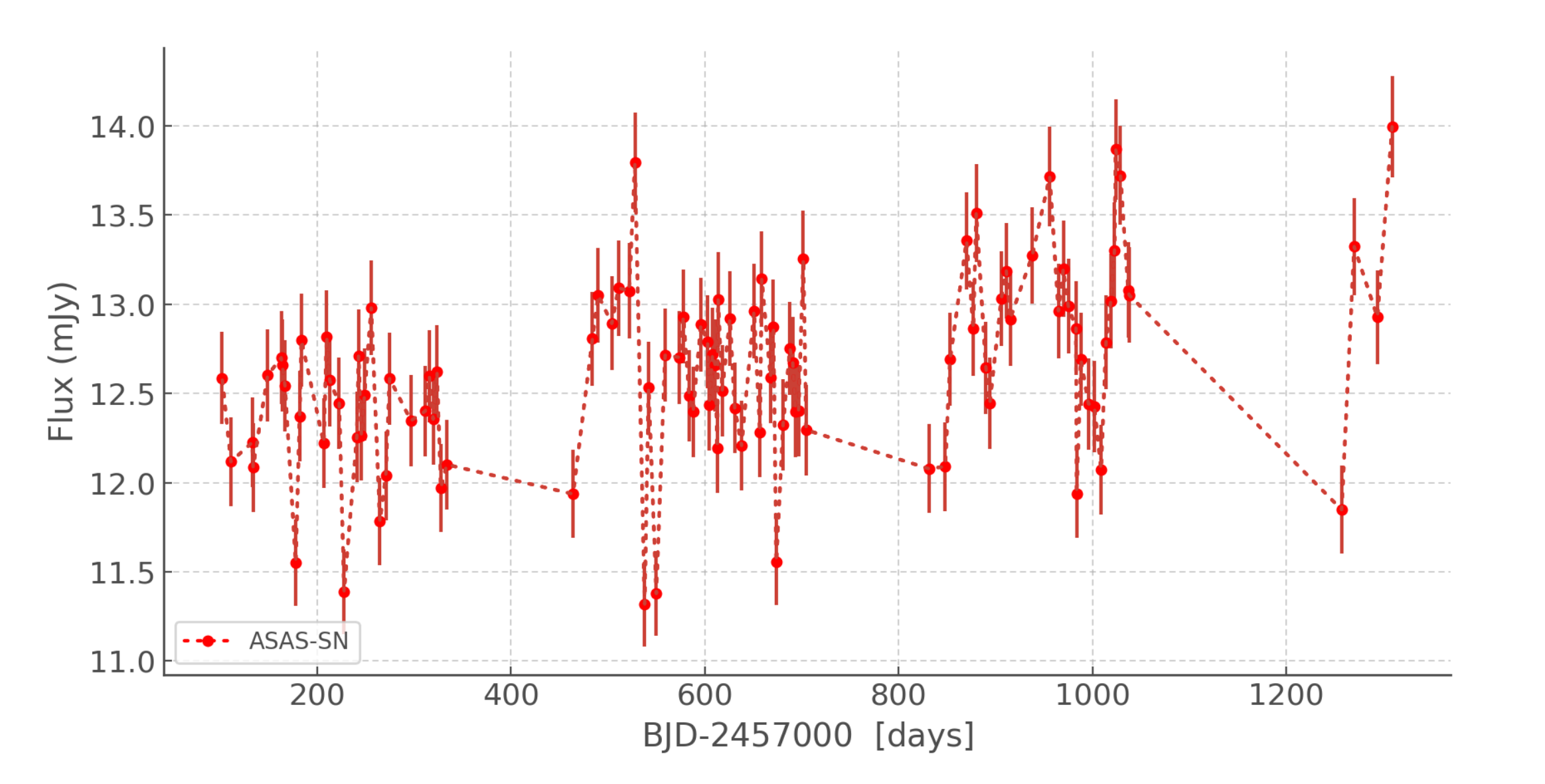}
   \includegraphics[scale=0.35,angle=270]{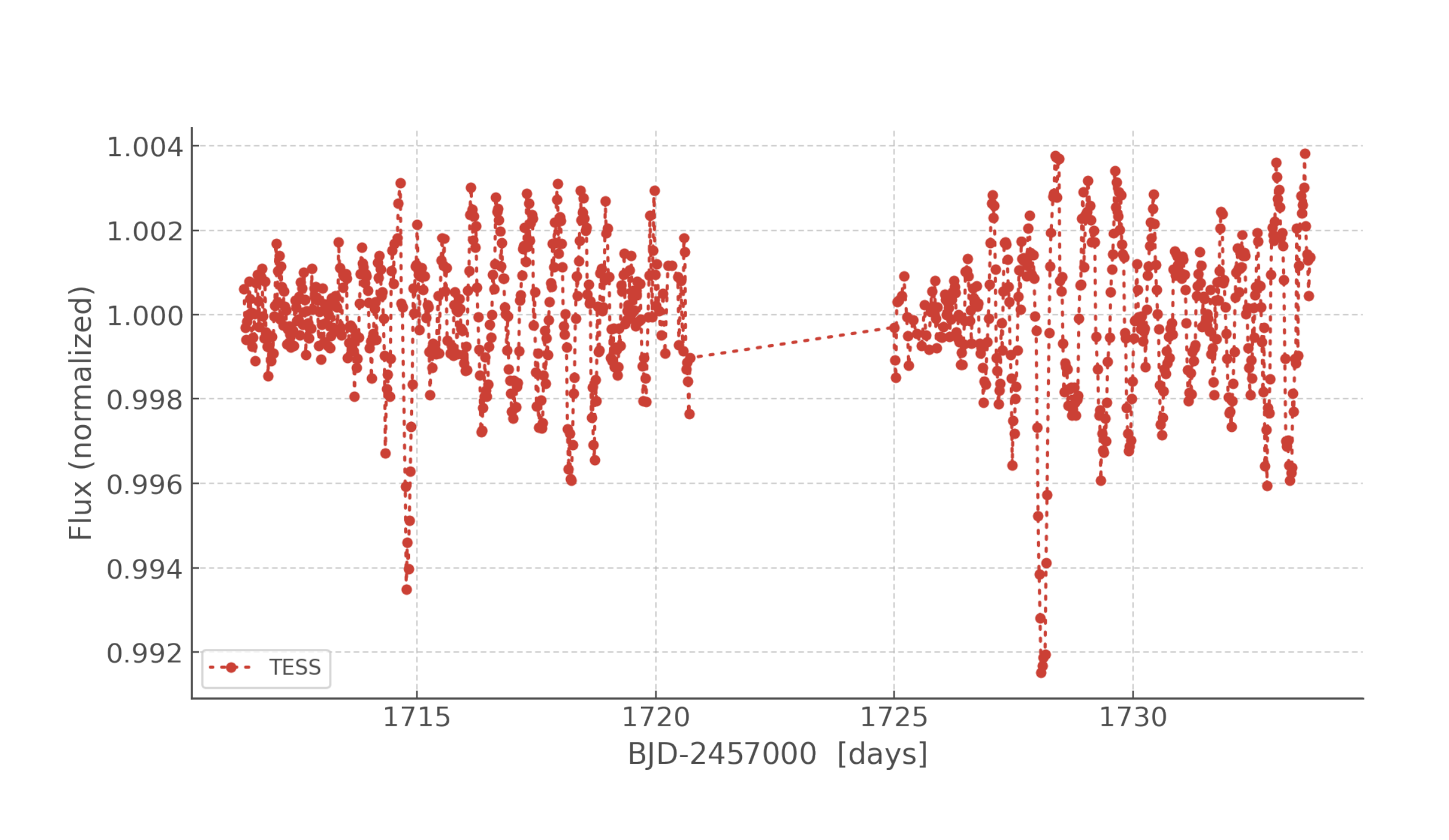}
   \caption{ASAS-SN (top) and TESS (bottom) light-curves for J20395358+4222505. The variability is clear for both dataset even if the amplitude and time-domain differ. The TESS light-curve has been cleaned from bad data.
   }
         \label{lightcurve}
       \end{figure}

       Inspection of the light-curves by the ASAS-SN \citep{kochanek17,shappee14} and TESS \citep{ricker15} projects (see Fig.~\ref{lightcurve}) reveals variability on the scale of days and years. The ASAS-SN light-curve\footnote{https://asas-sn.osu.edu} shows a modest increase, of the order of 1 mJy ($\Delta V \sim 0.15$ mag), in the flux of J20395358+4222505 during 3 years. For the TESS ligth-curve, normalized and detrended by Kepler spline and available at MAST as a HLSP\footnote{https://archive.stsci.edu/hlsp/qlp} \citep{huang20}, the observed variability after cleaning the curve from bad data (skipping data with quality flag higher than 2048) is dominated by the variability at the level of $\pm$0.2$\%$ in time scales of about 0.5 days (much shorter than the rotational period of 19.4$\times$sin$i$), typical of OB supergiants \citep[see f.e. Sect. 3.6 in][]{burssens20}. Two slightly stronger peaks ($\approx$1$\%$) are separated by about 13 days. With these two peaks, if the variability would be of binary or rotational origin, we would be able to constrain the corresponding inclination, but present data are insufficient to clearly distinguish between the different possibilities (stochastic or pulsational variability, rotational modulation, binarity). 

\subsection{Wind properties}
\label{wind}

\begin{figure}
   \centering
   \includegraphics[scale=0.5,angle=0]{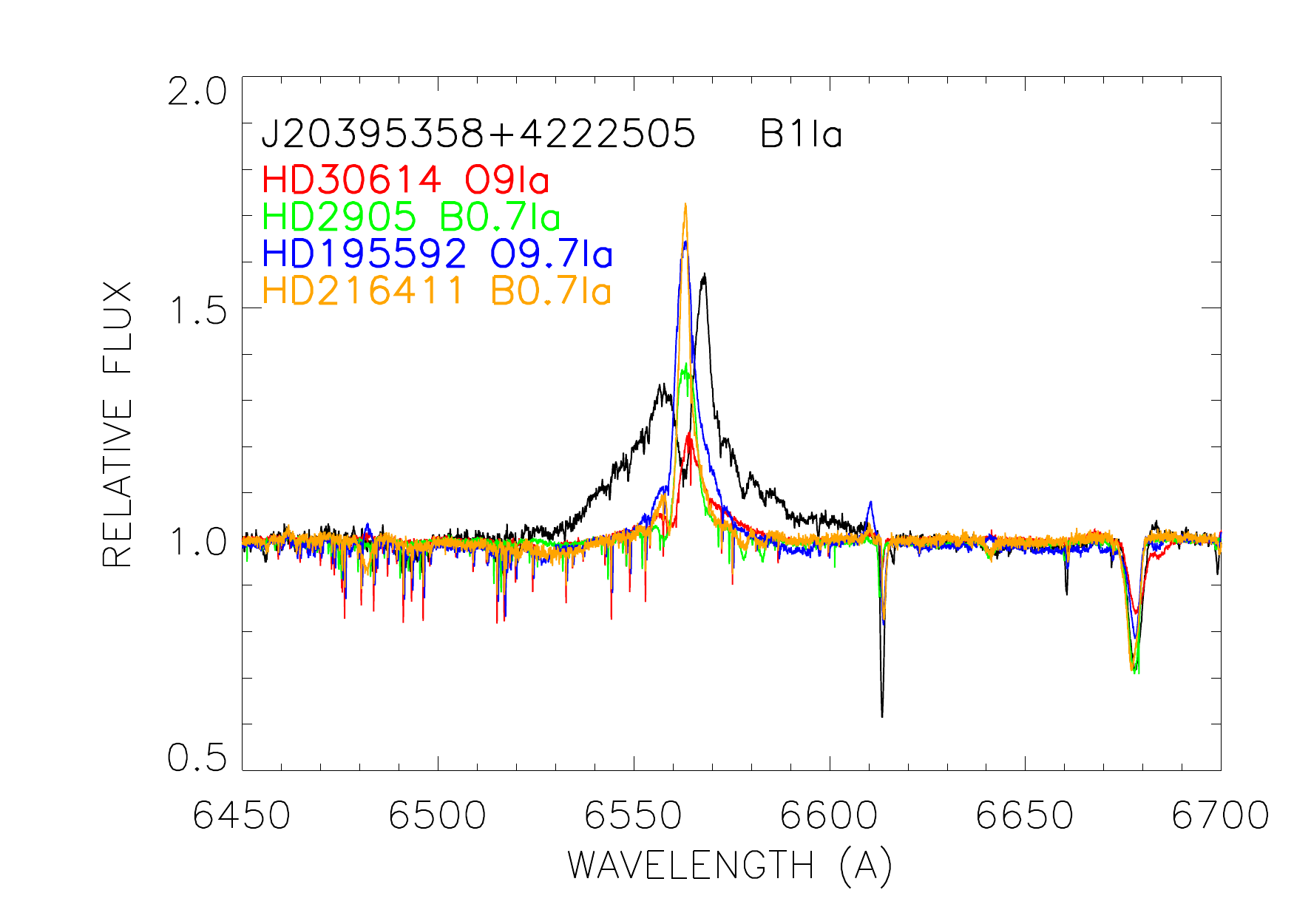}
      \caption{The H$_\alpha$ profile of \bhg~ compared with some of the strongest profiles in O9-B1 supergiants.  Other stars of similar spectral type with strong H$_\alpha$ emission (e.g., HD104565, OC9.7 Iab, HD148686, B1Iaeqp, or HD190603, B1.5Ia$^+$) show the same behaviour as the ones shown here: their H$_\alpha$ profiles are narrower than that of \bhg~ and the emission peaks are usually (not always) lower. The list in the left upper corner follows the height of the profile peaks. }
         \label{halpha}
   \end{figure}

The H$_\alpha$ emission of \bhg~ is exceptional for its spectral type, as the comparison with HD\,2905 in Fig.~\ref{spec_comp} has shown. In Fig.~\ref{halpha} we compare the  H$_\alpha$ profile from \bhg~ with those of the stars showing the strongest emission in a list of 176 Galactic B-supergiants currently under analysis (de Burgos et al., in prep.; the sample includes some O9 supergiants).  None of the inspected stars displays such a strong H$_\alpha$ profile, although in some cases the emission peak may be higher than that of \bhg. Even taking into account the variability of this line in B-supergiants \citep[see e.g.][]{ssimon18, haucke18} it is clear that \bhg~ shows a remarkable H$_\alpha$ profile.

The wind has a high terminal velocity as compared to stars of the same spectral type (see f.e. Fig. 11 in \citealt{clark12}, where the highest terminal velocity of a B1 supergiant is 1275 \kms~ for the peculiar star HD\,190066, a value taken from  \citealt{searle08}). Its ratio to the escape velocity (\vinf/$v_{escape}$= 2.7)\footnote{With $v_{escape}$ corrected for the (1-$\Gamma_e$) factor, with $\Gamma_e$ the ratio of radiation pressure acceleration due to electron scattering to gravitational acceleration} confirms that the star is
still at the hot side of the first bi-stability jump (BSJ), very close to the intermediate zone between the hot and cool sides (see Fig. 12 in \citealt{markova08}, also \citealt{petrov16}). The high \vsini~ also supports this, as B-supergiants on the cool side show lower projected rotational velocities.

The mass-loss rate derived from the stellar parameters we have obtained is 2.4$^{+0.2}_{-0.3}  \times$10$^{-6}$ \Msun a$^{-1}$, with values that range between 1.8 and 3.6$\times$10$^{-6}$ \Msun a$^{-1}$ when considering our lower and upper distance limits  (the unclumped values would amount to 5.7, 7.6 and 11.4$\times$10$^{-6}$ \Msun a$^{-1}$). Comparison with Fig.11 of \cite{clark12} shows how strong is the wind of \bhg. This can also be appreciated when comparing with other works in the literature. \cite{benaglia07} show that two stars in the \Teff= 20-25 kK range show an unusual large wind efficiency\footnote{The wind efficiency is defined as $\eta= {{\dot{M} v_\infty}\over{(L/c)}}$}: HD\,2905 and HD\,169454 (the latter one of the hypergiants in the sample of \cite{clark12}). The wind efficiency we obtain for \bhg~ ($\eta=$0.33) is comparable to that of those stars in \cite{benaglia07} ($\eta= $0.4-0.5), and higher if we consider the presence of clumping in our analysis. 

\begin{figure}
   \centering
   \includegraphics[scale=0.5,angle=0]{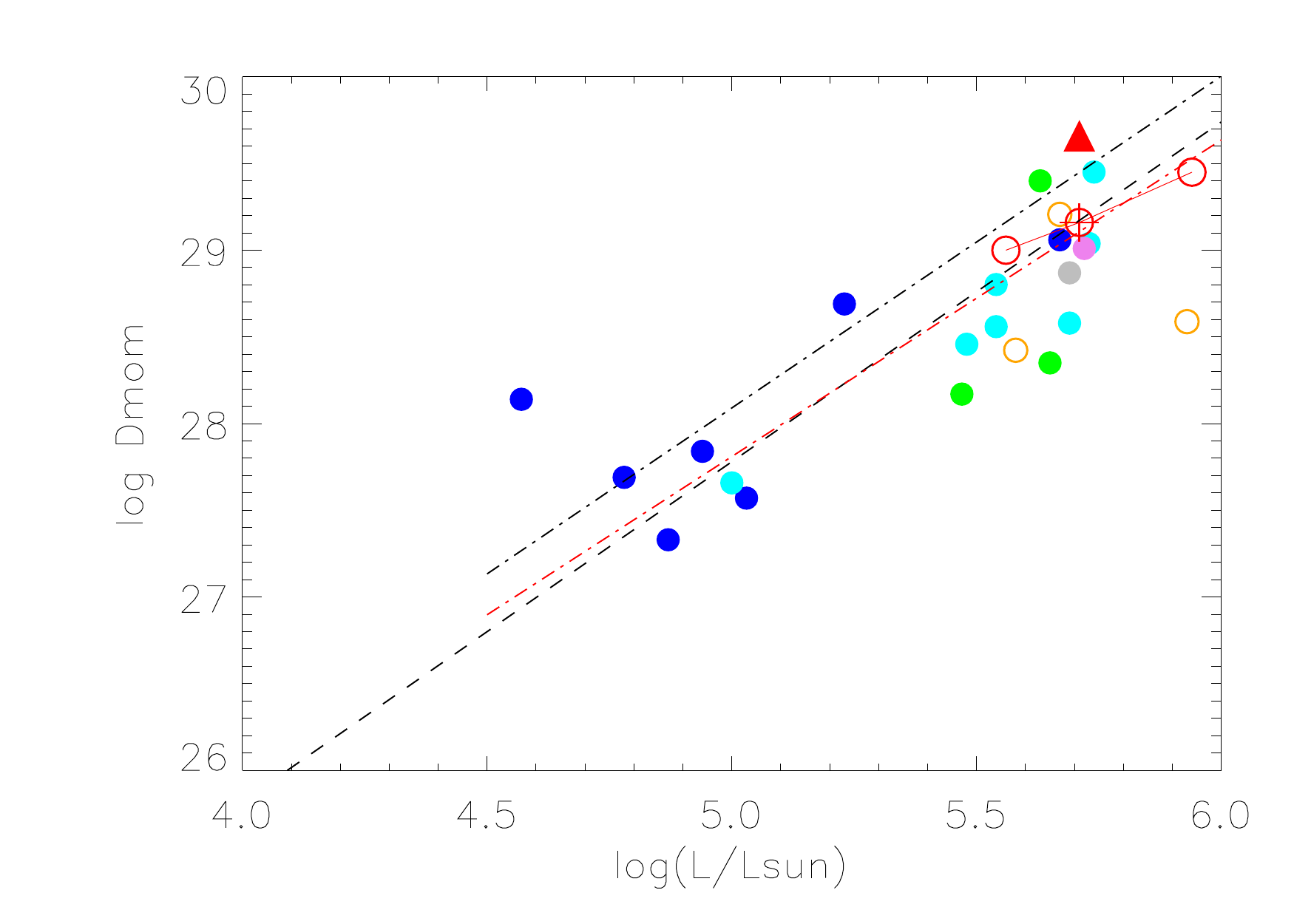}
   \caption{The Modified Wind Momentum--Luminosity Relationship for B0-B1.5 Galactic supergiants. Solid symbols represent unclumped values, hollow symbols are for values that include clumping. Data are from \protect\cite{crowther06} (green circles, it includes also the O9.5Ia star HD\,30614), \protect\cite{searle08} (cyan), \protect\cite{markova08} (magenta), \protect\cite{clark12}(orange, hollow circles), \protect\cite{haucke18} (blue) and \protect\cite{ssimon18} (grey). We also plot the values we obtain for \bhg~ at the different luminosities (hollow red circles) and the unclumped value we obtain for our star at the Cyg OB2 distance (solid red triangle). The cross in the central hollow symbol of \bhg~ gives the size of the error bars. For our star we have joined the clumped values corresponding to the different distances. The dashed line is the fit obtained by \protect\cite{haucke18} to their points and the dash-dotted lines are the relationships obtained by \protect\cite{vink00} for Galactic O (red line) and B (black line) stars.}
         \label{mwm}
   \end{figure}

In Fig.~\ref{mwm} we plot the Modified Wind-Momentum (D$_{mom}$)--Luminosity Relationship (WLR, \citealt{kudritzki95}) for the B0-B1.5 stars analyzed by different authors and compare it with the one we obtain for \bhg~ at the adopted distance of 1.76 kpc (the dot with the error bars) and at the lower and upper distance limits of 1.5 and 2.3 kpc. The plot includes values for Galactic B-supergiants by \cite{crowther06, searle08, markova08, clark12, haucke18} and \cite{ssimon18}. Only Clark et al. include clumping as we have done and thus we must be careful with the comparisons. For comparison, we have added the position of our star at the distance of Cyg OB2 when we do not consider clumping (i.e., multiplying by $\sqrt{10}$) and that of the O9.5Ia star HD\,30614 from \cite{crowther06}. Clumped and unclumped values are given by hollow and solid symbols, respectively. We also show the fit obtained by \cite{haucke18} to the early B-types (B0-B1.5) in their sample and the WLR predicted by \cite{vink00} for Galactic O and B stars (actually, stars at the hot and cool sides of the BSJ).

We see that the modified wind momentum of \bhg~ agrees well at any distance with the WLR relationship expected for stars at the hot side of the BSJ and with the relationship obtained by \cite{haucke18} for their early B-supergiants. But we note that we have included clumping. The modified wind momentum of \bhg~ is also comparable to (or larger than) many {\it unclumped} values for B-star in the literature. From the values by \cite{clark12} for the early-B hypergiants, only BP Cru (Wray 15-977, also a High-Mass X-Ray Binary) has a comparable modified wind momentum, precisely the star for which  \cite{clark12} obtain a clumping factor of f$_{cl}$=1. In contrast, the point representing the {\it unclumped} value of \bhg~ lies above any other point, including that of HD\,30614, and above the WLR for stars at the cool side of the BSJ. Thus, if our adopted clumping is typical for these stars, \bhg~ has an exceptionally high modified wind momentum and the typical early B-supergiants would be below the predicted values.

The actual mass-loss rate value of \bhg~ depends of course on the actual clumping structure in the wind, that we have just adopted here. Observations at longer wavelengths will be needed to get some information on its clumping, but it is clear that the wind of the star is stronger (i.e., with more momentum) than we would expect from its spectral classification. The H$_\alpha$ profile resembles that of much earlier spectral types (like mid-O supergiants). We have considered the possibility that the star has an earlier spectral type. For example, we see some weak \ion{Si}{iv} and \ion{He}{ii} 4686\AA. However, these features are consistent with a B0.5-B1 classification together with a high luminosity. More important, all other spectral features are consistent with the classification (f.e., the ratio of \ion{He}{i} to \ion{Mg}{ii} or the \ion{Si}{} spectrum) and imply that the star cannot have an earlier spectral type. The main uncertainty about the strong wind in \bhg~ comes from the H$_\alpha$ variability of B supergiants. \cite{haucke18} analyze the variability of Galactic B-supergiants and find variations in the H$_\alpha$ profile of all of them, in some cases very significant ones. In the most extreme cases, these variations translate into a factor of 2.7 in the mass-loss rate. If we decrease our mass-loss rate for \bhg~ by such a factor (assuming that we have observed the star during a particularly strong H$_\alpha$ emission phase) we still will have a very strong wind, but not one departing so remarkably from the average behaviour. Clearly, a follow-up spectroscopic campaign is required to solve this issue. Meanwhile \bhg~ should be included among the stars with a very strong stellar wind (for its spectral classification).

   \subsection{Evolutionary status of \bhg}
   \label{evolstat}
  
\begin{figure}
   \centering
   \includegraphics[scale=0.5,angle=0]{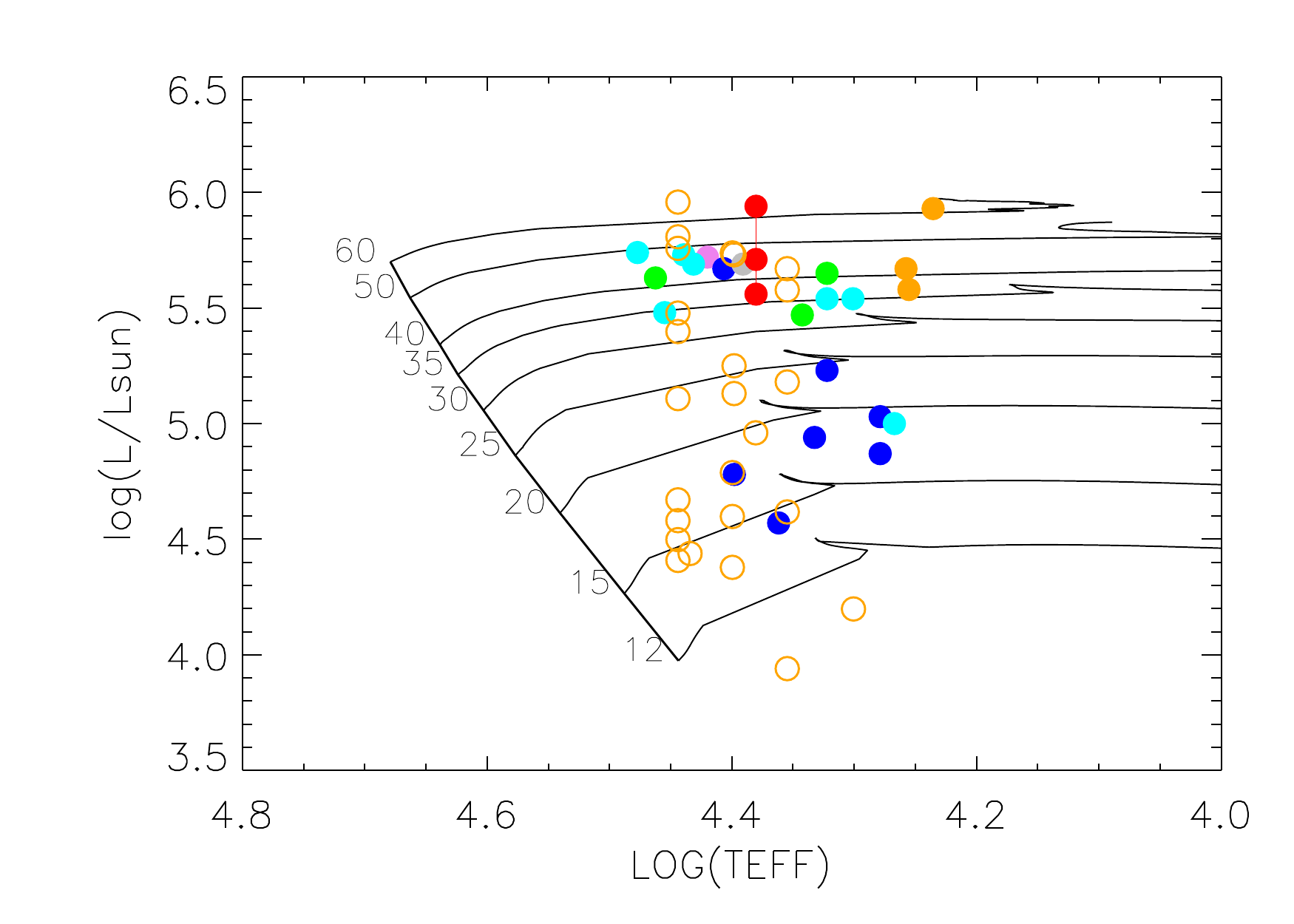}
      \caption{Hertzsprung-Russell Diagram with the position of \bhg~ for the adopted distance and the upper and lower distance limits adopted (solid red circles). The same stars as in Fig.~\ref{mwm} have been plotted, plus the B0-B1.5 supergiants in Cyg OB2, Cyg OB9 and their surroundings by \protect\cite{berlanas18} (open orange circles; these are calibrations, see text). The evolutionary tracks have been taken from \protect\cite{brott11} for stars with initial {\bf rotational velocities of 220} \kms. The numbers close to the Zero Age Main Sequence is the stellar initial mass.}
         \label{hrd}
       \end{figure}

   Figure~\ref{hrd} shows the Hertzsprung-Russell Diagram (HRD) with evolutionary tracks from \cite{brott11} for an initial rotational velocity of 220 \kms. We have chosen this value because, at the position \bhg~ in the HRD, it gives a rotational velocity close to the one observed ($\approx$100 \kms from the tracks versus \vsini=110 \kms observed; we note however, that the actual rotational velocity could be larger, as we do not know the inclination of the rotational axis see Sect.~\ref{photvar}). The position of \bhg~ in the diagram (adopted distance and upper and lower limits) is marked by the red solid circles. The same stars as in Fig.~\ref{mwm} with the same colour codes have also been plotted and we have added  the B0-B1.5 supergiants in Cyg OB2, Cyg OB9 and surroundings listed by \cite{berlanas18} (open orange circles).  We note that these latter authors use calibrations to get temperatures and luminosities from the spectral classification, not individual analyses. We have added them because they are in the same region as \bhg~ and probably belong to the same star-forming region (and possibly to the same association). Note also that \cite{berlanas18} used a distance modulus of 10.8 magnitudes for Cyg OB2, while we are using 11.2. We then increased their luminosities by 0.16 dex for the stars in Cyg OB2 and their surroundings for the comparison.
   
   We find that \bhg~ is a very luminous blue supergiant, comparable to the most luminous of its class and to some of the cooler blue hypergiants, a group that it will possibly join in the course of its evolution. The strong H$_\alpha$ emission, the observed variability and the high luminosity indicate that the star is possibly approaching such a phase. 
   As in the discussion in the previous section, we plot in Fig.~\ref{hrd} the points corresponding to the distances of 1.5 and 2.3 kpc. At the distance of 2.3 kpc the luminosity of \bhg~ would be \logl= 5.93, placing the star above any other at similar temperature, consistently with the strong H$_\alpha$ emission observed (we remember that H$_\alpha$ variability, or a combination of luminosity and variability, could also be an explanation for the strong H$_\alpha$ emission). Even at a distance as close as 1.5 kpc the star would have \logl= 5.55.  The tracks indicate an initial stellar mass around 46.0 \Msun, with a present mass of 39.9 \Msun, slightly lower but consistent with the spectroscopic value we obtained (46.5$\pm$15.0 \Msun; note that the mass discrepancy will in this case increase with mass, i.e., with increasing distance).  The mass-loss rate predicted by the tracks at the position of \bhg~ is 9.1$\times$10$^{-6}$ \Msun a$^{-1}$. This is very close to the one we have determined if we neglect the effect of clumping (7.6$\times$10$^{-6}$ \Msun a$^{-1}$). That means that the star should have lost less mass than predicted by the tracks if the effect of clumping remains approximately constant from the ZAMS up to the stage of \bhg, reducing the difference between the evolutionary and spectroscopic masses. As indicated before, this large mass-loss rate cannot be acting during a long period, except if we assume that the angular momentum is efficiently transported from the interior to the surface (as it is predicted by Brott et al. models).

   The star does not show a CNO pattern as could be expected because of the relatively advanced evolutionary stage. Only a mild C defficiency, based in only one line, is observed, while the \ion{N}{} abundance is found solar, with the models from \cite{brott11} predicting nearly twice this value (8.05 versus 7.83$\pm$0.15). The \ion{He}{} abundance also does not show indications of enrichment, and the other $\alpha$ elements analyzed, \ion{Si}{} and \ion{Mg}{} also point to solar abundances. Therefore, in spite of the advanced evolutionary phase and the high rotational velocity (and the possible efficient transport of angular momentum from the interior) we do not see evidence of a significant surface enrichment by Hydrogen burning products. These facts may be consistent with the evolution of a binary system in which \bhg~ would have been the initial secondary \citep{langer20}, gaining mass and angular momentum during the mass transfer, but without significant enrichment (a situation that would depend on the characteristics of the binary systems, like the initial period and mass ratio, and the details of the mass transfer process). This scenario would be supported by the possible radial velocity variations indicated by the B-band spectrum, but they require confirmation. We note however that the lack of evidence of a runaway nature of \bhg~ increases the chance that, if it formed initially a binary system with a more massive companion, the system is still bounded.

       \section{Conclusions}
       We have analyzed \bhg, a previously known supergiant (initially classified as B0 I) in the neighbourhood of Cyg OB2 and in the direction of the star-forming region DR21. Its high extinction along the line of sight made it inconspicous and concealed its high luminosity. Having realized it \citep{berlanas18} we observed the star during a MEGARA@GTC Commissioning run.
       
       We have found a difference in radial velocity between the spectrum taken in the first night of observation and the other spectra taken the night after. We find a small discrepancy between the position of the interstellar lines between those nights, but much smaller than the detected stellar radial velocity variations.

       We classify the star as B1 Ia based mainly on its \ion{Si}{} spectrum, the emission in H$_\alpha$ and the weakness of the higher Balmer emission lines and of the \ion{N}{ii} spectrum. The lack of a P-Cygni profile in H$_\beta$ prevents us from classifying it as Ia$^+$, and we suggest that this purely spectroscopic criterion should be adopted to distinguish classes Ia and Ia$^+$. The spectrum shows a very strong H$_\alpha$ emission profile for its spectral type, even when comparing with other early-B Ia supergiants or Ia$^+$ hypergiants.

       The analysis of the spectrum indicates a hot B supergiant with an unusually strong wind (for which we have assumed the clumping factor and characteristics based on the work by \citealt{clark12}), a high projected rotational velocity and solar abundances for all elements, except for a mild \ion{C}{} underabundance (based only on the \ion{C}{ii} 4267 line) and a possible slightly \ion{He}{} enrichment (but still compatible with solar). In particular, \ion{N}{} is found to be fully consistent with solar. Thus, in spite of the large rotational velocity, the abundance pattern seems to be free of significant enrichment.

       The analysis of the star is hampered  by the uncertainty in the distance to the star. We discarded the distance derived from Gaia DR2 parallax, and adopted the distance to DR21 (1.5 kpc) as a lower limit and the distance derived from eDR3 as an upper limit. The star is relatively isolated at a separation of 1.5$\degree$ from the centre of Cyg OB2, which at the distance of this association means a projected separation of about 45 pc. It has a proper motion that is consistent with the Cyg OB2 members. We adopt the distance to Cyg OB2 as the distance to the star, although we note that the connection is not strong and further work is required to probe it.
       
       Using that distance we derive the radius from the infrared photometry, and then mass, luminosity and mass-loss rate for that same distance. \bhg~ has a very high mass-loss rate for its spectral type of 2.4$^{+0.20}_{-0.30}$ $\times$10$^{-6}$ \Msun a$^{-1}$, where we have considered a clumping factor of f$_{cl}$=10. The star has a high wind terminal velocity that places it in the hot side of the bi-stability jump. We see that it agrees well with the expected WLR at this hot side. Compared to similar stars in the literature, however, its modified wind momentum is very large if our adopted clumping is typical of early B supergiants, which would mean that the theoretical value is too large for typical B-supergiants.

       We confirm that \bhg~ is a very luminous (\logl= 5.71$\pm$0.04) blue supergiant and we speculate that the star will join the group of B hypergiants in the near future (astronomically speaking). We obtain a spectroscopic mass of 46.5$\pm$15.0 \Msun. This is consistent with the evolutionary tracks by \cite{brott11}, that give an initial mass of 46 \Msun~ and a present mass of 39.9 \Msun. The difference between the spectroscopic and evolutionary masses could be reduced if the mass-loss rate adopted by the evolutionary computations (close to the one we obtain neglecting clumping) would be lower.

       The high rotational velocity indicates that the high mass-loss rate should be recent, or that the angular momentum transport from the interior is very efficient. However, this last possibility is not consistent with the lack of a clear CNO process pattern at the surface. These facts could also be explained with a binary scenario in which \bhg~ would be the initial secondary, assuming that the initial system had the required properties. The presence of a former more massive companion would be consistent with the lack of a peculiar proper motion in \bhg~ and the hints of radial velocity variation found in this work. But with only one spectrum supporting it, stronger evidence is required to confirm the presence of such putative companion.

       \section{acknowledgements}
       We would like to dedicate this paper to Simon Clark, our close friend whose inspiration and contribution to the advance on the field of Luminous Blue Variables and early hypergiants has been fundamental.

       We want to thank the referee, Dr. Ian Howarth, for his careful reading of the manuscript and the very useful report that helped improve this paper and correct mistakes.

       We also want to acknowledge fruitful discussions with N. Langer, I. Negueruela, P. Blay and R. Dorda. We thank A. de Burgos for the list of B-supergiants and their spectra.
       
S.S.-D. and A.H. acknowledge support from the Spanish Government Ministerio de Ciencia e Innovación through grants PGC-2018-091 3741-B-C22 and CEX2019-000920-S and from the Canarian Agency for Research, Innovation and Information Society (ACIISI), of the Canary Islands Government, and the European Regional Development Fund (ERDF), under grant with reference ProID2020010016.  M.G. and F. N. acknowledge financial support through Spanish grant PID2019-105552RB-C41 (MINECO/MCIU/AEI/FEDER) and from the Spanish State Research Agency (AEI) through the Unidad de Excelencia “María de Maeztu”-Centro de Astrobiología (CSIC-INTA) project No. MDM-2017-0737. S.R.B. acknowledges support by  the  Spanish  Government under grants AYA2015-68012-C2-2-P and PGC2018-093741-B-C21/C22 (MICIU/AEI/FEDER, UE). S.R.A. acknowledges funding support from the FONDECYT Iniciaci\'on project 11171025 and the FONDECYT Regular project 1201490. JIP acknowledges finantial support from projects Estallidos6 AYA2016-79724-C4 (Spanish Ministerio de Economia y Competitividad), Estallidos7 PID2019-107408GB-C44 (Spanish Ministerio de Ciencia e Innovacion), grant P18-FR-2664 (Junta de Andalucía), and grant SEV-2017-0709 “Centro de Excelencia Severo Ochoa Program” (Spanish Science Ministry). AGP, SP, AG-M, JG and NC acknowledge support from the Spanish MCI through project RTI2018-096188-B-I00.

This work has made use of data from the European Space Agency (ESA) mission
{\it Gaia} (\url{https://www.cosmos.esa.int/gaia}), processed by the {\it Gaia}
Data Processing and Analysis Consortium (DPAC,
\url{https://www.cosmos.esa.int/web/gaia/dpac/consortium}). Funding for the DPAC
has been provided by national institutions, in particular the institutions
participating in the {\it Gaia} Multilateral Agreement.

This research has made use of the VizieR catalogue access tool, CDS, Strasbourg, France (DOI: 10.26093/cds/vizier) and the VizieR Photometry Viewer Tool developed by Anne-Camille Simon and Thomas Boch. The original description of the VizieR service was published in Ochsenbein, Bauer \& Marcout, 2000, A\&AS 143, 23.

This publication makes use of data products from the Two Micron All Sky Survey, which is a joint project of the University of Massachusetts and the Infrared Processing and Analysis Center/California Institute of Technology, funded by the National Aeronautics and Space Administration and the National Science Foundation.

This publication makes use of data products from the Wide-field Infrared Survey Explorer, which is a joint project of the University of California, Los Angeles, and the Jet Propulsion Laboratory/California Institute of Technology, funded by the National Aeronautics and Space Administration.

This publication also makes use of data from the Spitzer, TESS, AKARI and HST satellites, the Carlsberg Meridian telescope at the ORM, the Pan-STARRS telecopes in Hawaii and the NOMAD, APASS and UCAC catalogs, as well as from the IACOB spectroscopic database.

\section{Data availability}
The data underlying this article will be shared on reasonable request to the corresponding author

\end{document}